\documentclass[11pt]{article}
\usepackage{amsfonts}
\usepackage{amsmath}
\newcommand{\ud}{\mathrm{d}}
\begin{document}
\begin{titlepage}
\begin{flushright}
CP3-07-14\\
ICMPA-MPA/2007/15\\
May 2007\\
\end{flushright}
\begin{centering}
 
{\ }\vspace{1cm}
 
{\Large\bf Topologically Massive Gauge Theories and}

\vspace{5pt}

{\Large\bf their Dual Factorised Gauge Invariant Formulation}

\vspace{0.7cm}

Bruno Bertrand

\vspace{0.3cm}

{\em Center for Particle Physics and Phenomenology (CP3)}\\
{\em Institut de Physique Nucl\'eaire}\\
{\em D\'epartement de Physique, Universit\'e catholique de Louvain (U.C.L.)}\\
{\em 2, Chemin du Cyclotron, B-1348 Louvain-la-Neuve, Belgium}\\
{\em E-mail: {\tt Bruno.Bertrand@fynu.ucl.ac.be}}

\vspace{0.8cm}

Jan Govaerts\footnote{On sabbatical
leave from the Center for Particle Physics and Phenomenology (CP3),
Institut de Physique Nucl\'eaire, Universit\'e catholique de Louvain (U.C.L.),
2, Chemin du Cyclotron, B-1348 Louvain-la-Neuve, Belgium,
E-mail: {\tt Jan.Govaerts@fynu.ucl.ac.be}.}

\vspace{0.3cm}

{\em Institute of Theoretical Physics}\\
{\em Department of Physics, University of Stellenbosch}\\
{\em Stellenbosch 7600, Republic of South Africa}\\

\vspace{0.3cm}

{\em UNESCO International Chair in Mathematical Physics and 
Applications (ICMPA)}\\
{\em University of Abomey-Calavi}\\
{\em 072 B.P. 50, Cotonou, Republic of Benin}\\

\vspace{0.7cm}

\begin{abstract}
\noindent
There exists a well-known duality between the Maxwell-Chern-Simons theory and the ``self-dual'' massive model in 2+1 dimensions. This dual description has been extended to topologically massive gauge theories (TMGT) in any dimension. This Letter introduces an unconventional approach to the construction of this type of duality through a reparametrisation of the ``master'' theory action. The dual action thereby obtained preserves the same gauge symmetry structure as the original theory. Furthermore, the dual action is factorised into a propagating sector of massive gauge invariant variables and a sector with gauge variant variables defining a pure topological field theory. Combining results obtained within the Lagrangian and Hamiltonian formulations, a new completed structure for a gauge invariant dual factorisation of TMGT is thus achieved.

\end{abstract}

\vspace{10pt}

\end{centering} 

\vspace{125pt}

\end{titlepage}

\setcounter{footnote}{0}

\section{Introduction}

A manifest realisation of the gauge invariance principle implies that the original fields used to define any gauge theory do not generate physical configurations since these fields are not gauge invariant degrees of freedom. As a matter of fact, two general approaches to isolate genuine physical degrees of freedom are available. The first involves some gauge fixing procedure in order to effectively remove the contributions of redundant gauge variant degrees of freedom. However such gauge fixings usually suffer Gribov problems, except in some exceptional cases. The second new approach, to be discussed presently, has been called ``dual factorisation''. Duality turns out to be especially interesting when an explicit correspondence between the fields of the two descriptions is achieved while at the same time exchanging their strong and weak coupling regimes\footnote{The theories to be considered hereafter are free and do not display this characteristic.}. In this Letter an interesting feature arising from a gauge covariant dualisation procedure is introduced. Indeed, following a convenient linear and local redefinition of the gauge fields within the Lagrangian formulation, gauge variant degrees of freedom are decoupled from the physical gauge invariant ones, thus implying a factorised dual formulation. This redefinition corresponds to the Lorentz covariant extension of a canonical transformation of phase space preserving canonical commutation relations within the Hamiltonian formulation. The main difficulty arising for such a programme is the fairly rare existence of such reparametrisations while at the same time being local and conserving the number of degrees of freedom.

In this Letter, our dual factorisation technique is illustrated for topologically massive gauge theories (TMGT) in which a topological term preserving exact gauge invariance generates a mass gap. Indeed a new general structure for dual factorisation involving both the Lagrangian and Hamiltonian formulations may then be completed for these TMGT. The dual action possesses the same gauge symmetry structure as the original theory and is factorised into a propagating sector of massive physical variables and a decoupled sector with gauge variant variables defining a pure topological field theory (TFT, for a review see \cite{Birmingham:1991ty}). In this context, dual factorisation is then called ``Physical-Topological'' (PT) factorisation~\cite{Bertrand:2007ja}.

\section{Dual Factorisation of the MCS Theory}

The existence of the factorised dual formulation is first shown in a simple example of topological mass generation, the well-known Maxwell-Chern-Simons (MCS) theory in 2+1 dimensions~\cite{MCS}. The general ``dual factorisation" advocated in the present Letter has already been developed in this specific three dimensional case~\cite{Guimaraes:2006gj} and within the Lagrangian formulation only as a particular application of the dual projection method developed originally within the context of the soldering formalism (for a review, see~\cite{Wotzasek:2003cw}).

A topological Chern-Simons term generates mass for a propagating spin one vector field $A$ of which the Lagrangian density reads, 
\begin{equation}\label{def:MCS_Lag}
\mathcal{L}_\textrm{MCS} = -\frac{1}{4\, e^2} F_{\mu\nu}\, F^{\mu\nu} + \frac{1}{4}\kappa\, \epsilon^{\mu\nu\rho}\, F_{\mu\nu}\, A_\rho  ,
\end{equation}
where the field scaling parameter $e$ is real while $\kappa$ is a real multiplicative constant. The reduction of a ``master'' Lagrangian~\cite{Deser:1984kw} accounts for the common origin of both the MCS and ``self-dual'' Lagrangians~\cite{Townsend:1983xs}. The master Lagrangian is the first order form of the MCS Lagrangian after the introduction of gauge invariant auxiliary fields $f_\mu$, readily reducible through Gaussian integration,
\begin{equation}\label{MCS_par}
\mathcal{L}^{2+1}_\textrm{master} = \frac{1}{2}e^2\, f_{\mu}\, f^{\mu} + \frac{1}{4}\epsilon^{\mu\nu\rho}\, F_{\mu\nu}\, \left(2\, f_\rho + \kappa\, A_\rho\right) \, .
\end{equation}
However, to a certain extent the reduction of the master Lagrangian as introduced in~\cite{Deser:1984kw} is analogous to a procedure of gauge fixing. Indeed the reduction of gauge variant variables within the Lagrangian formulation is analogous to the resolution of the associated first-class ``Gauss'' constraint within the Hamiltonian formulation.

In contradistinction to the master Lagrangian method~\cite{Deser:1984kw}, the dual factorised theory is constructed through a local and linear field redefinition, hence of field independent path integral Jacobian, leading to a redefinition of the master action $S_\textrm{master}[A,f] \to S_\textrm{SD}[E,\mathcal{A}]$, namely
\begin{eqnarray}\label{def:MCS_Transf}
E_\mu \left( A_\mu , f_\mu \right) &=& f_\mu , \nonumber\\
\mathcal{A}_\mu \left( A_\mu , f_\mu \right) &=& \frac{1}{\kappa} f_\mu + A_\mu .
\end{eqnarray}
Note that this field redefinition is well defined provided only the topological mass parameter $\kappa$ is non vanishing, $\kappa\ne 0$.
Upon reduction through Gaussian integration, the gauge invariant variables $E_\mu$ are found to correspond to the electric and magnetic field
components,
\begin{displaymath}
E_i \equiv \varepsilon_{ij}\, E^j_{\rm{elec}}, \qquad E_0 \equiv B_{\rm{mag}} .
\end{displaymath}
Consequently, a coherent reparametrisation of configuration space is achieved. In fact, it factorises the action into two decoupled contributions,
\begin{displaymath}
\mathcal{L}^{2+1}_\textrm{fact} = \mathcal{L}_\textrm{SD}\left[E_\mu, \partial_\mu E_\nu\right] + \mathcal{L}_\textrm{CS}\left[\mathcal{A}_\mu, \partial_\mu \mathcal{A}_\nu\right] .
\end{displaymath}
In deriving this expression a total surface term mixing the two field variables has been ignored, since it does not contribute for any appropriate choice of boundary conditions. It may, however, play a role when the quantum field theory is defined on a manifold with boundaries. The transformation (\ref{def:MCS_Transf}) resulting from the Lorentz covariant extension of the phase space canonical transformation introduced in \cite{Bertrand:2007ja} is a simpler though equivalent change of variable as compared to that used in~\cite{Guimaraes:2006gj}.

The physical self-dual part $\mathcal{L}_\textrm{SD}$ consists of Proca and topological mass terms, 
\begin{displaymath}
\mathcal{L}_\textrm{SD} = \frac{1}{2}e^2\, E_{\mu}\, E^{\mu} - \frac{1}{2\kappa}\epsilon^{\mu\nu\rho}\, \partial_\mu E_\nu\, E_\rho .
\end{displaymath}
This part describes a single propagating spin one free excitation of mass $m=\hbar\, \kappa\, e^2$ and violates parity.
The second part $\mathcal{L}_\textrm{CS}$ consists of gauge variant variables defining a purely topological Chern-Simons theory,
\begin{displaymath}
\mathcal{L}_\textrm{CS} = \frac{1}{2}\kappa\, \epsilon^{\mu\nu\rho}\, \partial_\mu \mathcal{A}_\nu\, \mathcal{A}_\rho .
\end{displaymath}
This last part, already expected within the path integral quantisation approach~\cite{Arias:1994xm}, is absent from the dual Lagrangian when the master action method~\cite{Deser:1984kw} is used in which case all the topological content inherited from the original Chern-Simons term is lost. In particular, non trivial topological features become manifest in the presence of external sources, or when the space manifold $\Sigma$ has non trivial topology (see~\cite{Dunne:1998qy} and references therein).

As far as the local part of the theory is concerned, the fact that the pure Chern-Simons theory describes gauge fields of flat connection implies, in combination with (\ref{def:MCS_Transf}), that
\begin{displaymath}
   \kappa\, \varepsilon^{\mu\nu\rho}\, \partial_\mu \mathcal{A_\nu} = \kappa\, \varepsilon^{\mu\nu\rho}\, \partial_\mu A_\nu + \varepsilon^{\mu\nu\rho}\, \partial_\mu f_\nu \approx 0 .
\end{displaymath}
One recovers of course the condition for the reduction of the master action in \cite{Deser:1984kw}, but in the present approach this condition is required as a weak constraint preserving the gauge content between the original and dual formulations.

\section{Dual Factorisation of General TMGT}

The equivalence between gauge non-invariant first order mass generating theories for any $p$-form and topologically massive gauge theories (TMGT) has so far been shown in diverse dimensions through the Hamiltonian embedding due to Batalin, Fradkin and Tyutin (BFT), either partial~\cite{Partial_BFT} or complete~\cite{Complete_BFT}, through the covariant gauge embedding method~\cite{Anacleto:2001rp,Menezes:2003vz} within the Lagrangian formulation, through the master action~\cite{Cantcheff:2001ws}, etc. All methods developed so far share a common characteristic, namely that in fact the dual action does not possess the same gauge symmetry content as the original formulation. Hence at the quantum level the equivalence between the two dual formulations applies only for pure theories defined on space manifolds of trivial topology.

The dual factorisation approach of this Letter readily applies to topological mass generation in any dimension and for all tensorial ranks. Given a real valued $p$-form field $A$ in $\Omega^p(\mathcal{M})$ and a $(d-p)$-form field $B$ in $\Omega^{d-p}(\mathcal{M})$ over a ($d+1$) dimensional spacetime manifold $\mathcal{M}$ endowed with a Lorentzian metric structure, the general action for TMGT reads
\begin{eqnarray} \label{def:TMGT_Action}
S[A,B] & = & \int_{\mathcal{M}} \frac{\sigma^{p}}{2 \, e^2}\, F\wedge\ast F + \frac{\sigma^{d-p}}{2 \, g^2} \, H\wedge\ast H \nonumber\\ 
& & + \kappa \int_{\mathcal{M}} (1-\xi) \, F\wedge B - \sigma^p \, \xi \, A\wedge H \, ,
\end{eqnarray}
where $\sigma = (-1)$. The arbitrary real and dimensionless variable $\xi$ introduced in order to parametrise any possible surface term is physically irrelevant for an appropriate choice of boundary conditions on $\mathcal{M}$. The field scaling parameters $e$ and $g$ are real. The action (\ref{def:TMGT_Action}) is invariant under two independent classes of finite abelian gauge transformations acting separately on either the $A$- or $B$-fields,
\begin{equation} \label{def:BF_gauge}
A'= A + \alpha , \qquad  B' =  B + \beta ,
\end{equation}
where $\alpha$ and $\beta$ are, respectively, closed $p$- and $(d-p)$-forms, while the derived quantities $F=\ud A$ and $H=\ud B$ are the gauge invariant field strengths of $A$ and $B$, respectively.
The last term in (\ref{def:TMGT_Action}) is a topological ``$BF$'' coupling between the two dynamical fields $A$ and $B$. In 3+1 dimensions, one recovers the Cremmer-Scherk action~\cite{TMGT}.

In order to construct the dual factorised action of TMGT, the original action (\ref{def:TMGT_Action}) must be written in its first order form after the introduction of gauge invariant auxiliary $(d-p)$- and $p$-form fields $\mathfrak{f}$ and $\mathfrak{h}$, respectively,
\begin{eqnarray}\label{def:TMGT_Parent}
 S_\textrm{master} & = & \frac{e^2}{2} \left( \mathfrak{f} \right)^2 + \frac{g^2}{2} \left( \mathfrak{h} \right)^2 + \int_{\mathcal{M}} F \wedge \mathfrak{f} + H \wedge \mathfrak{h} \nonumber\\
 & + & \kappa \int_{\mathcal{M}} (1-\xi) \, F\wedge B - \sigma^p \, \xi \, A\wedge H \, .
\end{eqnarray}
In (\ref{def:TMGT_Parent}) the inner product on $\Omega^k(\mathcal{M})\times\Omega^k(\mathcal{M})$ is defined as
\begin{displaymath}
     \left(\omega_k,\eta_k\right) = \int_{\mathcal{M}} \omega_k\wedge\ast\eta_k  ,
\end{displaymath}
with the convenient notation $(\omega_k)^2 = \sigma^{d+1-k}\, \left(\omega_k,\omega_k\right)$.
A simple local and linear transformation in the master action (\ref{def:TMGT_Parent}), of field independent path integral Jacobian and inducing the redefinition $S_\textrm{master}[A,B,\mathfrak{f},\mathfrak{h}] \to S_\textrm{fact}[E,G,\mathcal{A},\mathcal{B}]$, namely,
\begin{eqnarray}\label{def:TMGT_Transf}
E = \mathfrak{f}, &\quad& \mathcal{A} = A - \frac{1}{\kappa} \sigma^{p(d-p)}\, \mathfrak{h} , \nonumber\\
G = \mathfrak{h}, &\quad& \mathcal{B} = B + \frac{1}{\kappa} \mathfrak{f} ,
\end{eqnarray}
enables the factorisation of the theory into two decoupled sectors,
\begin{equation}\label{def:TMGT_fac}
    S_\textrm{fact}[E,G,\mathcal{A},\mathcal{B}] = S_\textrm{dyn}[E,G] + S_{BF}[\mathcal{A},\mathcal{B}] .
\end{equation}
Once again this transformation is well defined provided the topological coupling $\kappa$ does not vanish.
The two total divergences mixing the variables $\mathcal{A}$ and $\mathcal{B}$ with $E$ and $G$, respectively, are again parametrised by $\xi$.

The first contribution $S_\textrm{dyn}[E,G]$ consisting of dynamical gauge invariant variables reads as
\begin{eqnarray}
 S_\textrm{dyn} & = & \frac{e^2}{2} \left( E \right)^2 + \frac{g^2}{2} \left( G \right)^2 \nonumber\\
& + & \frac{1}{\kappa} \int_{\mathcal{M}} \sigma^{d-p}\, \xi\, E \wedge \ud G - (1-\xi)\, \ud E \wedge G \nonumber .
\end{eqnarray}
The gauge independent ``self-dual" action generalised to any dimension of~\cite{Menezes:2003vz,Cantcheff:2001ws} is recovered. Depending on the value of the parameter $\xi$, the Proca action for a $p$- or a $(d-p)$-form field is then readily identified through Gaussian integration. Indeed, by setting $\xi=1$ and integrating out the then Gaussian auxiliary $(d-p)$-form field $E$, one derives the action of a $p$-form field $G$ of mass $m=\hbar\mu$, with $\mu=\kappa\, e\, g$. Alternatively one may also obtain the action of a $(d-p)$-form field $E$ of mass $m=\hbar\mu$, by fixing $\xi=0$ and eliminating the Gaussian $p$-form field $G$.

The second contribution $S_{BF}[\mathcal{A},\mathcal{B}]$ to the dual factorised action (\ref{def:TMGT_fac}) involves gauge variant variables transforming as follows under the original abelian gauge symmetries (\ref{def:BF_gauge}),
\begin{equation}
\mathcal{A}'= \mathcal{A} + \alpha , \qquad  \mathcal{B}' =  \mathcal{B} + \beta ,
\end{equation}
and defines in fact once again a pure topological field theory of the $BF$ type,
\begin{displaymath}
S_{BF} = \kappa \int_{\mathcal{M}} (1-\xi) \, \mathcal{F}\wedge \mathcal{B} - \sigma^p \, \xi \, \mathcal{A}\wedge \mathcal{H} ,
\end{displaymath}
where $\mathcal{F} = \ud\mathcal{A}$ and $\mathcal{H} = \ud\mathcal{B}$. This decoupled TFT sector thus insures that the gauge structure of the original theory is preserved through dual factorisation. Moreover, as in the MCS case, the presence of this topological term, so far hardly evoked in the literature for very particular types of TMGT~\cite{Arias:1996et}, has dramatic consequences. First, as described in \cite{Bertrand:2007ja} within the context of canonical quantisation, this term controls the degeneracy of the physical spectrum of the original TMGT through topological invariants of the space manifold when it is of non trivial topology. Second, this topological term could be of prime importance for theories where the $p$-form fields are connections coupled to extended objects carrying the associated relevant charges.

The transformation (\ref{def:TMGT_Transf}) is nothing other than the Lorentz covariant extension, in combination with the expressions for conjugate momenta, of the canonical transformation in the phase space of the original TMGT within their Hamiltonian formulation, as recently introduced in \cite{Bertrand:2007ja}. This covariant generalisation emphasizes the universal character of the dual factorisation method, whatever the formulation of the theory, hence leading to the following general and completed structure.

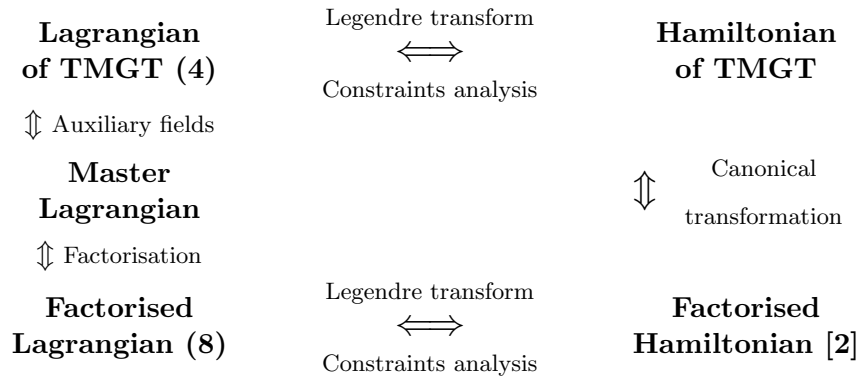
\begin{center}
\begin{tabular}{ccccc}
   \begin{tabular}{c}
   \textbf{Lagrangian} \\ \textbf{of TMGT (\ref{def:TMGT_Action})}
   \end{tabular}
& & 
   \begin{tabular}{c}
   \footnotesize{Legendre transform} \\ \Large{$\Longleftrightarrow$} \\ \footnotesize{Constraints analysis}
   \end{tabular}
& & 
   \begin{tabular}{c}
   \textbf{Hamiltonian} \\ \textbf{of TMGT}
   \end{tabular}
\\
$\Updownarrow$
   \footnotesize{Auxiliary fields}
& & \\
   \begin{tabular}{c}
   \textbf{Master} \\ \textbf{Lagrangian}
   \end{tabular}
& & & &
\Large{$\Updownarrow$}
   \begin{tabular}{c}
   \footnotesize{Canonical} \\ \footnotesize{transformation}
   \end{tabular}
\\
$\Updownarrow$
\footnotesize{Factorisation} &&&&\\
   \begin{tabular}{c}
   \textbf{Factorised} \\ \textbf{Lagrangian (\ref{def:TMGT_fac})}
   \end{tabular}
& & 
   \begin{tabular}{c}
   \footnotesize{Legendre transform} \\ \Large{$\Longleftrightarrow$} \\ \footnotesize{Constraints analysis}
   \end{tabular}
& & 
   \begin{tabular}{c}
   \textbf{Factorised} \\ \textbf{Hamiltonian~\cite{Bertrand:2007ja}}
   \end{tabular}
\end{tabular}
\end{center}

At first sight the introduction of the first order form of the action (\ref{def:TMGT_Action}) and thus the extension of the configuration space by auxiliary Gaussian fields seems artificial. As a matter of fact, to express directly the fields of the original Lagrangian formulation of TMGT as explicit functions of those of its dual formulation (\ref{def:TMGT_fac}) turns out to be impossible because the two formulations do not possess the same number of degrees of freedom. Although the two formulations describe the same physics, there are extra auxiliary degrees of freedom in the dual formulation. Therefore, a convenient Lagrangian must be chosen among those leading to the same constrained Hamiltonian~\cite{Govaerts:1991book}. The convenient formulation is the first order one (\ref{def:TMGT_Parent}) for which the comparison with the dual formulation is readily achieved from the local and linear transformation (\ref{def:TMGT_Transf}). This transformation simply redistributes the degrees of freedom, conserving the number of auxiliary fields and maintaining the gauge structure of the theory. In \cite{Bertrand:2007ja} where the dual Physical-Topological factorisation was achieved within the Hamiltonian formulation, all second-class constraints are being reduced using Dirac brackets. Therefore, the two phase spaces possess already the same number of degrees of freedom at any given spacetime point and dualisation is directly achieved. The first order form of TMGT makes manifest the relation between the covariant field redefinitions within the Lagrangian formulation and the associated canonical transformations within the Hamiltonian formulation.

\section{Conclusion}

The possibility of the dual factorisation introduced in this Letter is intimately related to the fact that TMGT generate a mass gap. Indeed within the Hamiltonian formulation this mass gap involves the non trivial dynamical global (or ``zero-mode'') sector (which carries the structure of harmonic oscillators). It is then possible to factorise phase space through a canonical transformation which is obviously local, using the mass gap parameter $\mu$ \cite{Bertrand:2007ja}. In this Letter this change of variables has been extended in a manifestly Lorentz covariant way by considering the first order form of the original Lagrangian of TMGT. In comparison to other methods developed so far in the literature, the technique consisting in constructing the dual action for TMGT by a local and linear redefinition of the fields is, firstly, much more direct and, secondly, preserves the gauge symmetry content of the original action, while at each step maintaining manifest Lorentz covariance. In this sense, this method enables to isolate the physical content of the theory in a gauge invariant way, the entire gauge variant contributions residing only in the second sector of the action which reduces to a pure topological field theory. The relevance of our conclusions for general TMGT is confirmed by some results already achieved for particular types of TMGT within the Lagrangian formulation alone or within the path integral framework~\cite{Guimaraes:2006gj,Arias:1994xm,Arias:1996et}.

The appearance of this topological sector which insures that the gauge symmetry content is maintained, has very intriguing consequences when TMGT are defined on topologically non trivial manifolds~\cite{Bertrand:2007ja} or are coupled to matter fields, whether of a fermionic or a bosonic character, since non trivial topological effects then arise. The coupling to matter fields is currently under investigation. One result of interest established so far is that in the symmetry breaking phase the effective abelian Maxwell-Higgs Lagrangian is equivalent to a particular form of TMGT coupled to a real scalar ``Higgs'' field in a very specific way~\cite{Bertrand:2007prep}.

Dual factorisation is the archetype of a more ambitious project whose basic ideas were suggested in a heuristic way in~\cite{Govaerts:2004bb}. If this kind of technique is to turn out to be applicable to large classes of gauge invariant theories generating a mass gap, it may offer perspectives in the development of new approximation schemes for non perturbative dynamics. In particular, it would be of great interest to understand if similar considerations could apply to matter fields coupled to Yang-Mills theories in order to isolate the low energy physical configurations of condensed matter states which reside in the zero-mode sector.

\section*{Acknowledgements}

The work of B.~B. is supported by a Ph.D. Fellowship of
the ``Fonds pour la formation \`a la Recherche dans l'Industrie et dans l'Agriculture'' (F.R.I.A.),
of the Associated Funds of the National Fund for Scientific Research (F.N.R.S., Belgium).

J.~G. acknowledges the Institute of Theoretical Physics for an Invited Research Staff position at the
University of Stellenbosch (Republic of South Africa).
He is most grateful to Profs. Hendrik Geyer and Frederik Scholtz, and the School of Physics
for their warm and generous hospitality during his sabbatical leave, and for financial support.
His stay in South Africa is also supported in part by the Belgian National
Fund for Scientific Research (F.N.R.S.) through a travel grant.

J.~G. acknowledges the Abdus Salam International Centre for Theoretical
Physics (ICTP, Trieste, Italy) Visiting Scholar Programme
in support of a Visiting Professorship at the UNESCO-ICMPA (Republic of Benin).

This work is also supported by the Institut Interuniversitaire des Sciences Nucl\'eaires and by
the Belgian Federal Office for Scientific, Technical and Cultural Affairs through
the Interuniversity Attraction Poles (IAP) P6/11.

\end{document}